\documentclass[12pt]{article}
\usepackage{graphicx}

\begin{document}

\newcommand{\hdblarrow}{H\makebox[0.9ex][l]{$\downdownarrows$}-}
\title{Soliton-like Spin State in the A-like Phase of $^3$He
in Anisotropic Aerogel}

\author{V.V. Dmitriev\footnote{Kapitza Institute for Physical Problems, 2
Kosygin Str., Moscow 119334, Russia}~~~D.A. Krasnikhin$^*$  N.
Mulders\footnote{Department of Physics and Astronomy, University
of Delaware, Newark, Delaware 19716, USA}~~~D.E. Zmeev$^*$}

\date{\today}

\maketitle

\begin{abstract}

We have found a new stable spin state in the A-like phase of
superfluid $^3$He confined to intrinsically anisotropic aerogel.
The state can be formed by radiofrequency excitation applied while
cooling through the superfluid transition temperature and its NMR
properties are different from the standard A-like phase obtained
in the limit of very small excitation. It is possible that this
new state is formed by textural domain walls pinned by aerogel.

PACS numbers: 67.57.Lm, 67.57.Pq
\end{abstract}

\section{Introduction}
Recently it was found\cite{anisotropy} that the anisotropy of
aerogel samples significantly affects the superfluid properties of
$^3$He in aerogel. In particular, NMR properties of the A-like
superfluid phase in anisotropic aerogel samples qualitatively
correspond to the properties of the superfluid A-phase of bulk
$^3$He with the order parameter vector {\bf l} fixed with respect
to the sample\cite{anisotropy,undisturbed}. One of the crucial
properties that point to the anisotropy of the sample is the
negative NMR shift in the A-like phase in certain orientations of
external static magnetic field {\bf H}. The analysis of data
obtained in superfluid $^3$He in aerogel over the last decade
shows that at least some of the aerogel samples are intrinsically
anisotropic, meaning that either the procedure of their growth
leads to different "porosity" in one of the directions (namely the
axis of cylindrical samples) or the samples get irreversibly
deformed at one of the stages of making the experimental cell. The
theoretical findings are that large enough anisotropy destroys the
state that would have existed in isotropic aerogel\cite{Volovik,
Fomin} and stabilizes a phase with the bulk A phase order
parameter corresponding to a spatially uniform
Anderson-Brinkman-Morel (ABM) order parameter.

Here we present results which show that in the A-like phase in
anisotropic aerogel apart from the spin state with properties
corresponding to the ABM order parameter a state with quite
different NMR properties can exist.

\section{Experimental details}

We have performed NMR experiments in two samples of 98.2\% open
silica aerogel in the superfluid A-like phase of $^3$He. Both
samples had a cylindrical shape. The first one was 4 mm in
diameter and 3.5 mm in height (sample 1), the other was 5 mm in
diameter and 1.5 mm in height (sample 2). The experimental cells
were made from Stycast-1266 epoxy and the walls of the cells did
not compress the samples at low temperatures when the epoxy
shrunk. In that sense the anisotropy (if any) of these aerogel
samples was intrinsic. Experiments were performed at pressures of
26.0 bar (sample 1) and 28.6 bar (sample 2) in magnetic fields
from 40 to 467 Oe (corresponding to NMR frequencies from 132 to
1517 kHz).

Apart from the longitudinal field solenoid (which produced a
static magnetic field oriented parallel to the axes of the
cylindrical samples~$\hat {\bf z}$) we had a coil that produced a
static field in the direction perpendicular to the longitudinal
field and to the direction of the RF (radio-frequency) excitation
field. It allowed us to rotate {\bf H} by any angle. The
homogeneity of the transverse external field was much worse than
that of the longitudinal field (10$^{-3}$ and 2$\times10^{-4}$
respectivly).

For longitudinal NMR experiments in sample 1 we had a high-Q
circuit which consisted of cold capacitors with teflon dielectric
(C=0.5 $\mu$F) and a superconducting NbTi coil (300 turns) with
its axis parallel to $\hat {\bf z}$. The resonant frequency of the
circuit was 9095 Hz, and the quality factor was 1860. In the
experiments with this sample we also used another NbTi coil for
transverse NMR (80 turns). We did not have the longitudinal NMR
circuit in the experiments with sample 2 and had a copper coil
($\sim$60 turns) for the transverse NMR.

We used a quartz tuning fork resonator for the
thermometry\cite{fork}. It was calibrated against a vibrating wire
resonator and proved very reliable.

\section{Results}

\begin{figure}[h]
\begin{center}
\includegraphics[scale=1.0]{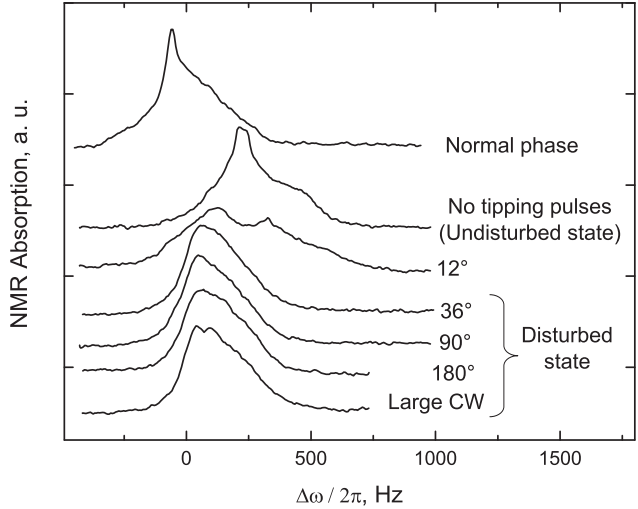}
\end{center}
\caption{CW NMR lines in the undisturbed and disturbed states in
transverse field in sample 1. The top line was observed above
T$_{ca}$, all other lines at T=0.89\,T$_{ca}$. The conditions in
the vicinity of T$_{ca}$ are indicated next to the lines, numbers
are amplitudes of the corresponding tipping pulses (see text). The
external field was perpendicular to the sample axis. For clarity
the zero levels of absorption are shifted. H=118 Oe, P=26.0 bar,
T$_{ca}$=0.80\,T$_c$.}\label{CW}
\end{figure}

The properties of the A-like phase were very similar in both
aerogel samples and no qualitative difference in the NMR
properties was found at different NMR frequencies for a given
orientation of {\bf H}. Continuous wave (CW) NMR measurements have
shown that the superfluid in our samples manifests properties
similar to those observed in an anisotropic aerogel sample
squeezed along its axis\cite{anisotropy}. In particular, in the
A-like phase the NMR shift in longitudinal field ({\bf
H}$\parallel{\hat {\bf z}}$) was negative and in transverse field
({\bf H}$\perp{\hat {\bf z}}$) it was positive. Thus we suggest
that in our samples the A-like phase corresponds to the A phase of
bulk $^3$He with the order parameter vector {\bf l} fixed with
respect to the aerogel sample\cite{undisturbed}. However, in both
samples we were able to create a novel spin state which is not
known for the bulk A phase: while cooling down in transverse field
through the superfluid transition temperature T$_{ca}$ in the
range $\sim$(1.02$\div$0.97)T$_{ca}$ we repeatedly applied tipping
pulses of a certain amplitude every few seconds for five to ten
minutes. Fig. \ref{CW} shows CW NMR lines obtained after
performing such a procedure for several tipping pulse amplitudes
compared to the NMR line obtained on cooling down without any
pulses. It is clear that the A-like phase enters different spin
states depending on whether or not the tipping pulses of
sufficiently large amplitude were applied in the mentioned
temperature range. We call the former the {\it disturbed} state
and the latter the {\it undisturbed} state. For the tipping angle
of 12$^\circ$ the disturbed state was formed only in part of the
sample. The disturbed state was also formed if a large enough
resonant continuous RF excitation ($\sim$0.01 Oe) was applied to
the sample while cooling down through T$_{ca}$. In the
longitudinal field the disturbed state did not form after any
tipping pulses or continuous excitation were applied in the
vicinity of T$_{ca}$. Also the NMR lines in either state did not
change after application of any RF excitation below
$\approx$\,0.95T$_{ca}$. The disturbed state proved to be stable:
no change in the NMR line was seen over a period of one day.

\begin{figure}[h]
\begin{center}
\includegraphics[scale=1.0]{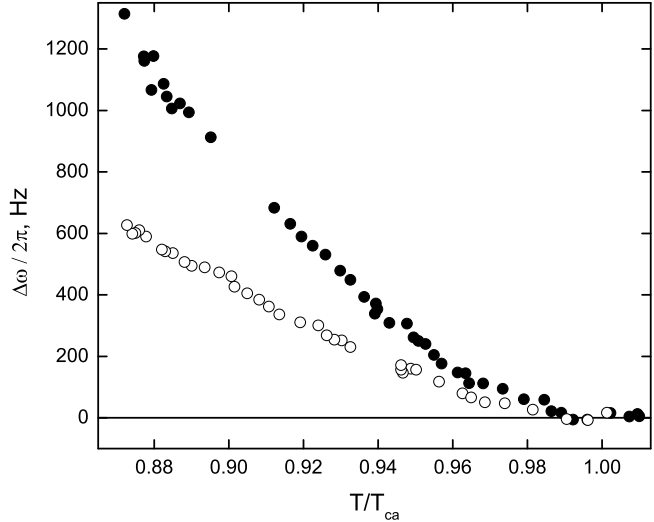}
\end{center}
\caption{Shifts of the mean resonant frequency in the disturbed
({\Large $\circ$} ) and undisturbed ({\Large $\bullet$}) states in
transverse field in sample 2. The shift in the disturbed state was
about 2 times smaller than in the undisturbed state. H=97.5 Oe,
P=28.6 bar. T$_{ca}$=0.82\,T$_c$.}\label{CW_shifts}
\end{figure}

The spin dynamics of the disturbed state turned out to be very
different from the dynamics of the undisturbed state. The most
evident difference was in the magnitude of the NMR frequency shift
from the Larmor value. In sample 2 the shift in the disturbed
state was 2 times smaller than in the undisturbed state (Fig.
\ref{CW_shifts}). We think that in sample 1 the anisotropy was not
spatially homogeneous because the width of the NMR line in the
A-like phase was large. Netherveless we were able to cool the
sample down to the A-like$\rightarrow$B transition and then warm
up so that the A-like phase survived only in part of the sample
and the other part was in the B-phase. In this case only the most
shifted part of the A-like phase signal (corresponding to the
region where {\bf l}$\parallel${$\hat {\bf z}$)
remained\cite{undisturbed}. The opposite phenomenon of survival of
the less shifted part was observed in\cite{anisotropy}. The A-like
phase NMR lines in both states obtained after such a procedure
were rather narrow and the ratio of the shifts in the disturbed
and undisturbed states was close to 3.

We were also able to slowly rotate {\bf H} when our sample was in
the disturbed state. The frequency shift of the NMR line became
negative in the longitudinal field but smaller (in absolute value)
than in the undisturbed state. No changes in the NMR line were
observed when the field was returned to the transverse
orientation. What is more surprising is that the disturbed state
was not modified when we decreased the magnetic field down to no
more than 0.5 Oe and returned the field to its former value.

In sample 1 we have also carried out longitudinal NMR experiments.
In these experiments we swept the temperature while recording the
signal from the longitudinal NMR coil. Because the axis of this
coil was oriented along $\hat {\bf z}$ we were not able to see any
signal for transverse orientation of {\bf H}. Therefore we used
angles $\psi = 0^{\circ}$ and $60^{\circ}$ between {\bf H} and
$\hat {\bf z}$. As is expected for the ABM order parameter no
longitudinal NMR was found in the undisturbed state for
$\psi=0^\circ$. For $\psi=60^\circ$ the longitudinal NMR signal in
the undisturbed state was clearly seen and its position on the
temperature axis was in a good agreement with that expected from
the value of the frequency shift measured by transverse
NMR\cite{undisturbed}. On the other hand for $\psi=60^\circ$ no
response in the disturbed state was observed in the experiments on
longitudinal resonance down to transition to the B-phase.

\begin{figure}[h]
\begin{center}
\includegraphics[scale=1.0]{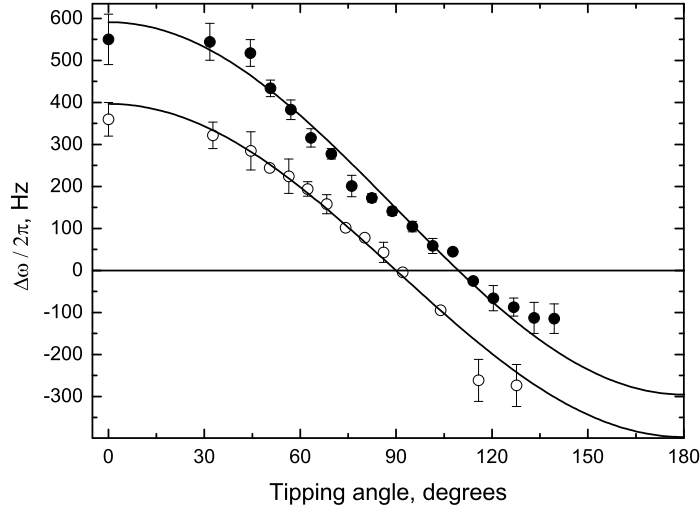}
\end{center}
\caption{Frequency shift as a function of the magnetization
tipping angle $\beta$ in pulsed NMR experiments in sample 2 in
transverse field. In the disturbed state ({\Large $\circ$}) the
shift fitted to $A_1\cdot\cos\beta$ with $A_1=395$ Hz and in the
undisturbed state ({\Large $\bullet$}) the shift fitted to
$A_2\cdot(1+3\cos\beta)/4$ with $A_2$=590 Hz. The points at zero
tipping angle are from CW NMR measurements.  H=97.3 Oe, P=28.6
bar. The temperature was 0.928T$_{ca}$ for the disturbed state and
0.933T$_{ca}$ for the undisturbed state.} \label{pulsed}
\end{figure}

One more difference between the two states is demonstrated by the
results of the pulsed NMR shown in Fig.\ref{pulsed}. We have found
that in the disturbed state the frequency shift depends on the
tipping angle $\beta$ as
$\Delta\omega_{dist}/2\pi=A_1\cdot\cos\beta$, while in the
undisturbed state the dependence was like in the ABM phase:
$\Delta\omega_{undist}/2\pi=A_2\cdot(1+3\cos\beta)/4$.

\section{Conclusions}

The properties of the disturbed state are very different from the
properties of the undisturbed state. First, the NMR shift is
several times smaller in the disturbed state. Second, no
longitudinal resonance was observed in this state. Third, the NMR
frequency shift depends on the tipping angle as
$\Delta\omega_{dist}/2\pi=A_1\cos\beta$.

The origin of the disturbed state is unclear. While the properties
of the undisturbed state can clearly be related to the A phase
order parameter with the vector {\bf l} fixed by anisotropy, the
properties of the disturbed state can not be explained in terms of
this order parameter in a straightforward manner. It is probable
that the disturbed state corresponds to textural domain walls
(``solitons") similar to those existing in the A phase of bulk
$^3$He\cite{solitons, soliton0, soliton1, soliton2}. It is known
that textural defects (at least vortices in the B-like phase) can
be strongly pinned by aerogel\cite{vort}. In our case solitons can
appear near T$_{ca}$ due to motion of the spin part of the order
parameter and at lower temperature can be pinned by aerogel, which
may explain their stability. However, not all properties of the
disturbed state can be explained by the soliton model. It is not
clear how tipping pulses as small as 12$^\circ$ can create domain
walls of the order parameter vectors: it is known that in the bulk
A phase solitons can be created only with much larger
pulses\cite{solitons, soliton0}. We should also note that in the
disturbed state the whole NMR line is changed while solitons in
the A phase either manifest themselves as small satellite peaks to
the main NMR line (i.e. solitons occupy only small part of the
sample) or are unstable (disappearing after a time of the order of
a minute). Also the difference in NMR shifts between the bulk
$^3$He-A with solitons and defectless superfluid is not as large
as in the case of the disturbed and undisturbed states in aerogel.
Note that tipping pulses altered the undisturbed state only in the
vicinity of T$_{ca}$ where the behavior of the A-like phase in
anisotropic aerogel is not described by the model of spatially
homogeneous A phase: as was found in\cite{undisturbed} the NMR
shift in the A-like phase at these temperatures is close to zero
and starts to grow only below $\sim$0.98\,T$_{ca}$. No RF
excitation can further influence the disturbed or undisturbed
states at lower temperatures, while in the bulk A phase solitons
can be created in a broad range of temperatures, down to 0.9T$_c$.

Properties of the so-called {\it c}-state observed earlier in
(presumably) isotropic the A-like phase\cite{c-state} are
qualitatively the same as the properties of the disturbed state.
The {\it c}-state was also formed by RF pulses applied near
T$_{ca}$ and its NMR line had also frequency shift a few times
smaller than the standard NMR line. However we can not confidently
identify the disturbed state with the {\it c}-state because the
value of the NMR shift in the {\it c}-state was typically about 5
times smaller than in the disturbed state (in samples 1 and 2) at
similar conditions, and in a longitudinal field the shift was
positive. We should note that in the sample used in\cite{c-state}
the shift of a standard A-like phase NMR line in longitudinal
field was also positive and its value was much smaller than in the
undisturbed state in the anisotropic aerogel. Presumably it was
due to the squeezing of the sample in transverse plane by spacers
on the side walls of the cell. This could decrease the intrinsic
anisotropy of the sample, so the A-like phase order parameter
could be different from the homogeneous A phase order
parameter\cite{undisturbed}. It also makes a direct comparison of
the disturbed state and the {\it c}-state impossible.

\section{Acknowledgements}

We thank I. Fomin and G. Volovik for useful discussions. The
research was supported by the Russian Foundation for Basic
Research (06-02-17185), the Ministry of Education and Science of
Russia (NSh-9725.2006.2) and CRDF (RUP1-MO-04-2632).

\end{document}